\documentclass[amsmath,amssymb,aps,prx,reprint,superscriptaddress]{revtex4-2}

\usepackage{amsmath}
\usepackage{xcolor}
\usepackage{amsfonts,amssymb}
\usepackage{graphicx}
\usepackage{hyperref}
\usepackage{academicons}
\usepackage{orcidlink}
\usepackage{multirow}
\usepackage{chemformula}

\begin{document}

\title{Overcoming systematic softening in universal machine learning interatomic potentials by fine-tuning}

\author{Bowen Deng\,\orcidlink{0000-0003-4085-381X}}
\affiliation{Department of Materials Science and Engineering, University of California, Berkeley, California 94720, United States}
\affiliation{Materials Sciences Division, Lawrence Berkeley National Laboratory, California 94720, United States}

\author{Yunyeong Choi\,\orcidlink{0000-0002-7946-4232}}
\affiliation{Department of Materials Science and Engineering, University of California, Berkeley, California 94720, United States}
\affiliation{Materials Sciences Division, Lawrence Berkeley National Laboratory, California 94720, United States}

\author{Peichen Zhong\,\orcidlink{0000-0003-1921-1628}}
\affiliation{Department of Materials Science and Engineering, University of California, Berkeley, California 94720, United States}
\affiliation{Materials Sciences Division, Lawrence Berkeley National Laboratory, California 94720, United States}

\author{Janosh Riebesell\,\orcidlink{0000-0001-5233-3462}}
\affiliation{Cavendish Laboratory, University of Cambridge, J. J. Thomson Ave, Cambridge, UK}

\author{Shashwat Anand\,\orcidlink{0000-0001-8805-1547}}
\affiliation{Materials Sciences Division, Lawrence Berkeley National Laboratory, California 94720, United States}

\author{Zhuohan Li\,\orcidlink{0000-0001-5372-9450}}
\affiliation{Materials Sciences Division, Lawrence Berkeley National Laboratory, California 94720, United States}

\author{KyuJung Jun\,\orcidlink{0000-0003-1974-028X}}
\affiliation{Department of Materials Science and Engineering, University of California, Berkeley, California 94720, United States}
\affiliation{Materials Sciences Division, Lawrence Berkeley National Laboratory, California 94720, United States}

\author{Kristin A. Persson\,\orcidlink{0000-0003-2495-5509}}
\affiliation{Department of Materials Science and Engineering, University of California, Berkeley, California 94720, United States}
\affiliation{Materials Sciences Division, Lawrence Berkeley National Laboratory, California 94720, United States}

\author{Gerbrand Ceder\,\orcidlink{0000-0001-9275-3605}}
\email[]{gceder@berkeley.edu}
\affiliation{Department of Materials Science and Engineering, University of California, Berkeley, California 94720, United States}
\affiliation{Materials Sciences Division, Lawrence Berkeley National Laboratory, California 94720, United States}

\date{\today}

\begin{abstract}
Machine learning interatomic potentials (MLIPs) have introduced a new paradigm for atomic simulations. Recent advancements have seen the emergence of universal MLIPs (uMLIPs) that are pre-trained on diverse materials datasets, providing opportunities for both ready-to-use universal force fields and robust foundations for downstream machine learning refinements. However, their performance in extrapolating to out-of-distribution complex atomic environments remains unclear. In this study, we highlight a consistent potential energy surface (PES) softening effect in three uMLIPs: M3GNet, CHGNet, and MACE-MP-0, which is characterized by energy and force under-prediction in a series of atomic-modeling benchmarks including surfaces, defects, solid-solution energetics, phonon vibration modes, ion migration barriers, and general high-energy states. 

We find that the PES softening behavior originates from a systematic underprediction error of the PES curvature, which derives from the biased sampling of near-equilibrium atomic arrangements in uMLIP pre-training datasets. We demonstrate that the PES softening issue can be effectively rectified by fine-tuning with a single additional data point. Our findings suggest that a considerable fraction of uMLIP errors are highly systematic, and can therefore be efficiently corrected. This result rationalizes the data-efficient fine-tuning performance boost commonly observed with foundational MLIPs. We argue for the importance of a comprehensive materials dataset with improved PES sampling for next-generation foundational MLIPs.

\end{abstract}


\maketitle

\section{Introduction}

\begin{figure*}[t]
\centering
\includegraphics[width=\linewidth]{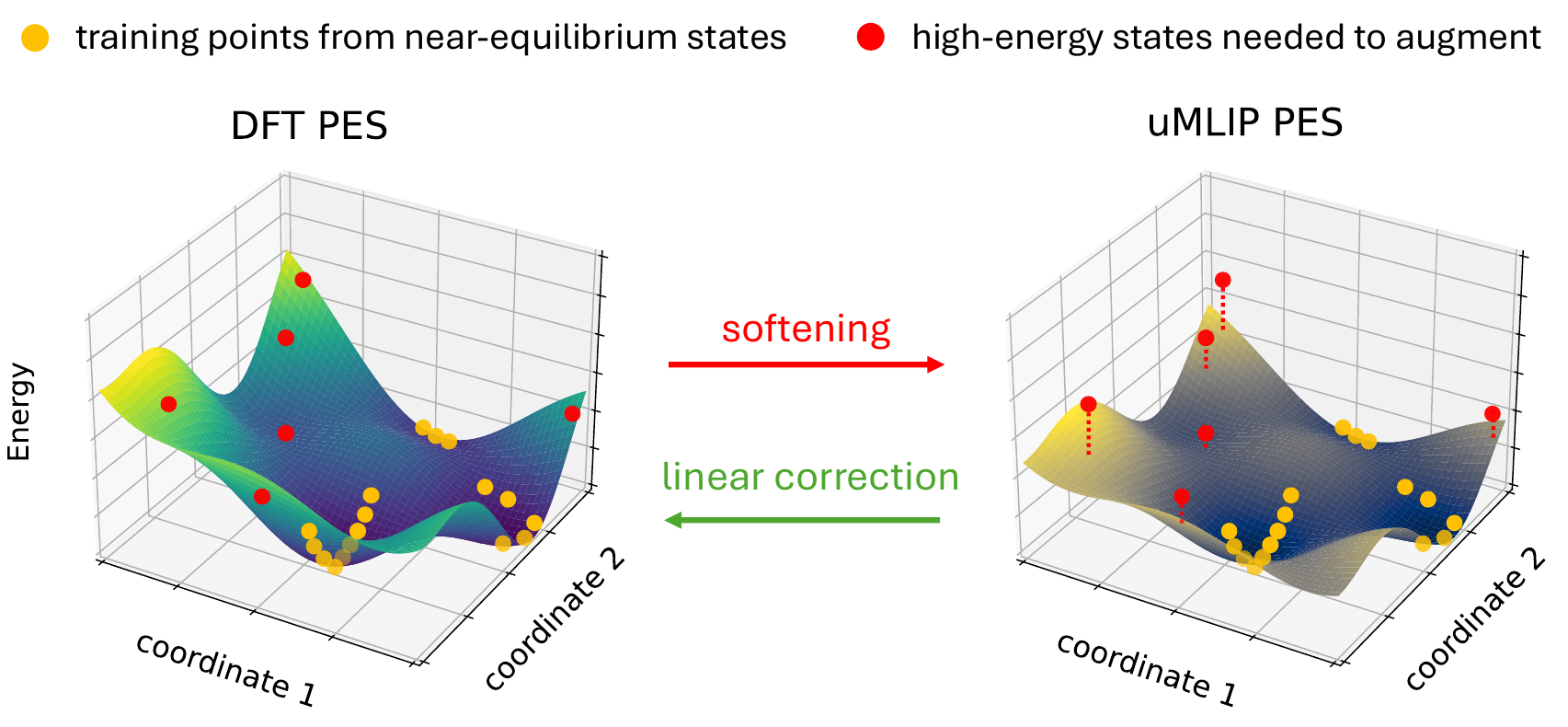}
\caption{\textbf{Potential energy surface softening in uMLIPs}. Left: schematic representation of the potential energy surface (PES) described in density functional theory (DFT), with two arbitrary coordinate axes. Right: PES described by universal machine learning interatomic potentials (uMLIPs), which well describes the PES regions sampled by near-equilibrium states in the pre-training dataset (orange), but experience larger errors in high-energy regions (red) with under-prediction of energies and forces. The softening behavior is largely systematic and, therefore can be efficiently fixed by a linear correction with a small amount of data augmentation.}
\label{fig:pes} 
\end{figure*}

Artificial intelligence (AI) is increasingly shifting the paradigm of scientific discovery to accelerate research and solve real-world scientific challenges \cite{AI_science_review_2023}. While \textit{ab-initio} quantum mechanical simulation methods, such as density functional theory (DFT), offer the theoretical foundation to investigate material and chemical science problems at the atomic scale, their computational demands limit their applicability in both spatial and temporal scales. Recent advancements in machine learning interatomic potentials (MLIPs) \cite{DPMD_2018, NequIP_2022} have enabled the opportunity to scale up quantum mechanical methods to million atoms simulations such as water, copper \cite{Guo_2022_billion}, and biomolecules \cite{Albert_2023_Biomolecular}. 

Alongside improvements in atomic environment descriptors and graph neural networks that enhance the expressivity of MLIP models \cite{DimeNet_2020, NequIP_2022}, universal machine learning interatomic potentials (uMLIPs) have demonstrated another avenue by taking advantage of pre-training on large and comprehensive material datasets \cite{M3GNet_2022, CHGNet_2023, Choudhary_2023_alignnff, MACEMP_2023, PFP_2022, GNOME_2023, Zhang_2024_DPA1}. These uMLIPs enable out-of-box atomic modeling covering the entire periodic table as well as providing robust machine-learning foundations for fine-tuning downstream tasks. While uMLIPs hold considerable promise, a critical challenge lies in their ability to reliably generalize to complex and diverse chemical environments, particularly those that deviate significantly from the pre-training data distribution. Several recent benchmark efforts have tested the uMLIPs' ability to identify stable materials \cite{Matbench_2023}, surface energies \cite{SurfaceBenchmark_2024}, lattice relaxations and vibrational properties \cite{Benchmark_uMLIP_2024}, etc. A systematic understanding of the ability of uMLIPs to extrapolate to common atomic-modeling tasks, especially those with atomic environments that are out of distribution (OOD), remains an open question with implications for their real-world applicability in material discovery and design.

In this work, we systematically investigate the extrapolative capabilities of three foundational uMLIPs -- M3GNet \cite{M3GNet_2022}, CHGNet \cite{CHGNet_2023}, and MACE-MP-0 \cite{MACEMP_2023} (hereafter referred to as MACE) -- across a diverse suite of OOD material modeling tasks, including surface energies, defect energies, solid-solution energetics, phonon vibrational modes, ion migration barriers, and high-energy transition states. Across all benchmark tests for all uMLIP models, our analysis reveals a consistent potential energy surface (PES) softening behavior, characterized by a systematic underprediction of energies and forces, as illustrated in Fig. \ref{fig:pes}. We attribute the PES softening issue to the combination of the biased sampling of near-ground-state configurations in the uMLIP pre-training datasets \cite{MaterialProject}, which primarily comprise DFT ionic relaxation trajectories near PES local energy minima. The uMLIPs trained predominantly on small energy and force labels suffer from distribution shifts and experience increased but systematic prediction errors in high-energy PES regions which are important for the kinetics of rare events, such as ion migrations, and for the energy of defects with undercoordinated atoms, such as vacancies and surfaces. 

We demonstrate that this systematic PES softening can be effectively mitigated by fine-tuning with a minimal amount of data points. We find that a simple linear correction derived from a single DFT reference label is sufficient to remove much of the PES softening issue, significantly enhancing the performance and robustness of uMLIPs. We rationalize this observation by arguing that a considerable amount of prediction errors in pretrained uMLIPs are highly systematic, and therefore can be efficiently corrected by modifying a limited fraction of the model parameters with only a small amount of data augmentation. Our work provides a theoretical foundation for the widely observed data-efficient performance boosts achieved by fine-tuning uMLIPs and highlights the advantage of atomic modeling with large and comprehensive foundational AI models.

\section{Results}

\subsection{Machine Learning Interatomic Potentials Framework}

MLIPs approximate the total energy of a system as a sum of atomic contributions, each dependent on the positions and chemical identities of the atoms in their local environment:
\begin{equation}
E = \sum_i^n E_i(\{\vec{r}_i\}, \{C_i\}), \quad f_i = - \frac{\partial E}{\partial \vec{r}_i}.
\end{equation}
$E_i$ is a learnable function that maps the set of position vectors $\{\vec{r}_i\}$ and chemical species $\{C_i\}$ of the neighboring atoms to the energy contribution of atom $i$. The force $f_i$ acting on each atom is calculated as the derivative of the total energy with respect to its position. In the training process, the parameters of the MLIP model are optimized to minimize the discrepancy between the predicted energies and forces and the corresponding reference values from the DFT labels.

The design of the atomic environment descriptor function $E_i$ is crucial to developing accurate and efficient MLIPs. To capture the essential physics and chemistry of the system, $E_i$ should be informative and satisfy proper translational and rotational symmetries. This is typically achieved through the use of graph representations \cite{Xie_Grossman_2018}, high-order interactions \cite{DimeNet_2020, M3GNet_2022}, the preservation of SE(3)/E(3)-equivariance using tensor products based on spherical harmonics \cite{NequIP_2022, MACEMP_2023}, Fourier basis \cite{luo2024_Gaunt_tensor}, or Cartesian-coordinates-based atomic density expansion \cite{cheng2024cartesian}. Additionally, the incorporation of chemical information, such as charge \cite{Ko_Finkler_Goedecker_Behler_2023} or atomic magnetic moment \cite{CHGNet_2023}, has been shown to enhance the predictive power of MLIPs.

In addition, recent efforts have been made to pre-train MLIPs on large open-sourced materials datasets such as the Materials Project \cite{MaterialProject}, which primarily consists of DFT ionic relaxation trajectories of various compounds and elements across the periodic table. While initial benchmarks have shown the promising applicability of universal MLIPs in predicting bulk materials energetics \cite{Matbench_2023, Benchmark_uMLIP_2024}, their performance and limitations in OOD atomic configurations require more benchmarking as the energy of these configurations is often directly relevant for practical materials behavior. The following sections present a systematic assessment of the uMLIPs' ability to extrapolate to low-symmetry OOD atomic configurations that are crucial for atomic-modeling tasks.

\subsection{Surface Energies}

\begin{figure*}[t]
\centering
\includegraphics[width=\linewidth]{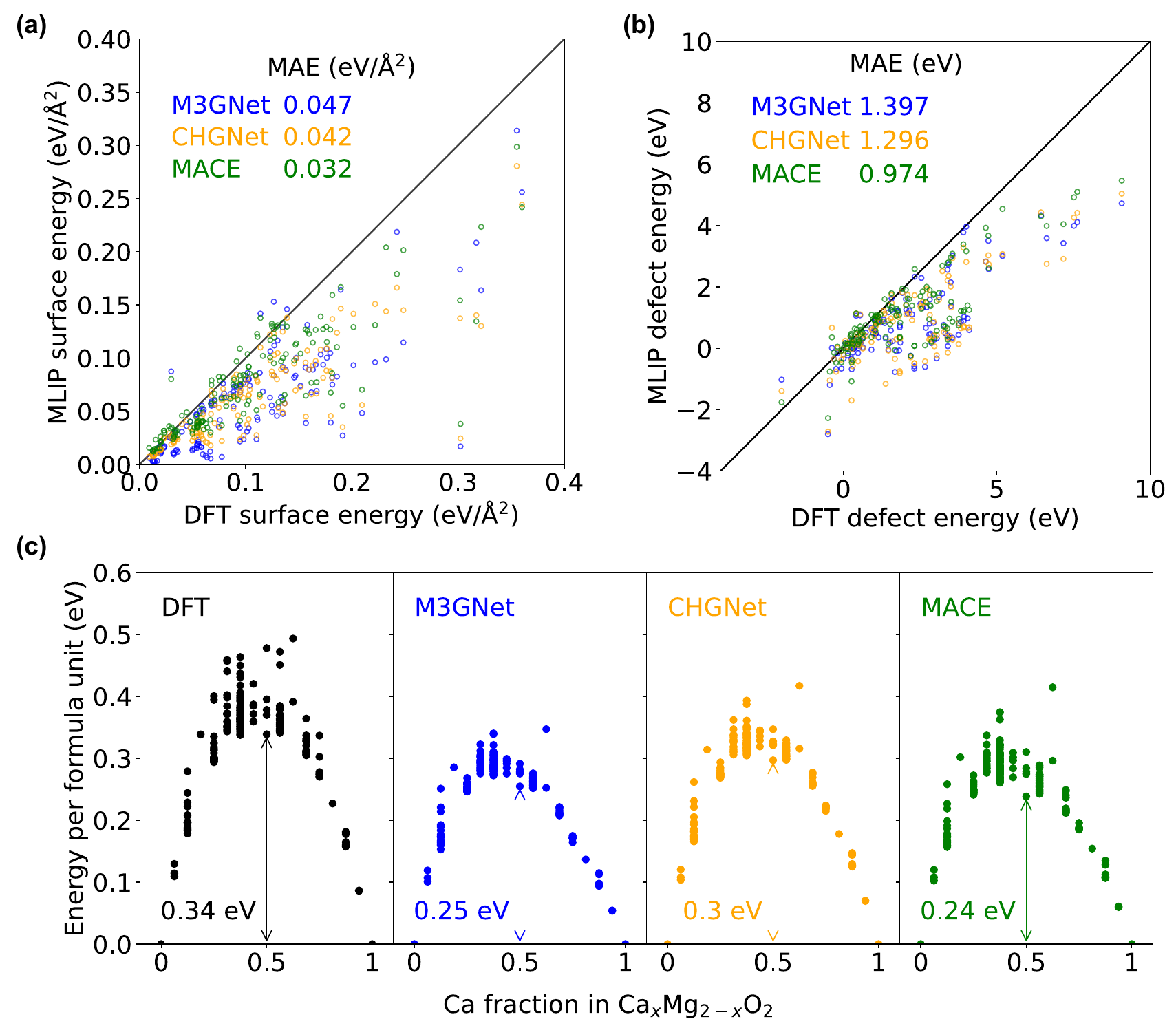}
\caption{\textbf{uMLIP performance on surfaces, defects, and solid solutions.} (a) Comparison of DFT surface energies and MLIP surface energies, evaluated on 147 surfaces from 29 chemical systems. (b) Comparison of DFT defect energies and MLIP defect energies, evaluated on 134 point defects from 32 chemical systems. (c) Formation energies in Ca$_x$Mg$_{2-x}$O$_2$ solid solution from DFT and uMLIPs. Each point corresponds to the energy of a specific Ca-Mg cation arrangement at a given Ca fraction. The distributions of energies are collectively underestimated, which would lead to an underprediction of the miscibility gap temperature in uMLIPs compared to DFT.}
\label{fig:surface_defect} 
\end{figure*}

Surface energies play an important role in determining the stability and morphology of materials, especially at the nano-scale where the surface-to-volume ratio is significant. Accurate prediction of surface energies is crucial for various applications such as catalysis \cite{OC20}, corrosion \cite{Corrosion_2015}, adhesion \cite{adhesion_2019}, nucleation \cite{Sun_2019_NC_nucleation}, and thin film growth \cite{ThinFilm_2000}. In this section, we assess uMLIP's performance in predicting surface energies, which are calculated as
\begin{equation}
    \gamma_{\text{surface}} = \frac{E_{\text{slab}} - E_{\text{bulk}}}{2A_{\text{slab}}},
\end{equation}
where $E_{\text{slab}}$/$E_{\text{bulk}}$ are the relaxed energies of the slab/bulk structures that can be obtained independently using DFT or MLIP methods in a large supercell approach. $A_{\text{slab}}$ denotes the surface area of the slab.

The energies of 147 surfaces with multiple Miller indices of 29 elements and binary compounds are evaluated, including Si, Cu, Al$_2$O$_3$, LiF, ZnS, etc. The DFT and uMLIP calculation details are listed in the \hyperref[sec:Methods]{Methods} section and \hyperref[sec:SI]{Supplementary} Table 1 lists the full set of elements and compounds with their corresponding prediction errors. Figure \ref{fig:surface_defect}(a) shows the uMLIP surface energies versus the DFT surface energies for the three uMLIPs tested, where MAE stands for the model's mean absolute error. MACE exhibits relatively better performance compared to CHGNet and M3GNet, achieving a MAE of 0.032 eV/$\mathrm{\AA}$. All three uMLIPs consistently underestimate the surface energies compared to DFT, except for a few predictions made by MACE and M3GNet. The trend in our result is consistent with the recent evaluation of \citet{SurfaceBenchmark_2024} on the surface energies of elemental crystals.

\subsection{Defect Energies}
We also analyze the accuracy of uMLIPs in calculating point defect energies, which is crucial for understanding a material's vacancy formation \cite{vacancy_2021}, dopabilities \cite{HT_defect_2023}, mechanical properties \cite{Defect_mechanical_2019}, and ionic mobilities \cite{Kang_2006_defect}. Specifically, we perform benchmarks for point defects including vacancies, interstitials, and anti-site defects. In metallic systems, the point defect energy can be calculated from the energy of a defect structure referenced to the corresponding perfect structure and the external chemical potential of the species added or removed
\begin{equation}
    E_i^\text{point defect} = E_i^\text{defect} - E^\text{bulk} - \Sigma \mu_i\Delta N_i,
\end{equation}
where $\mu_i$ is the chemical potential of the species \textit{i} forming the defect and $\Delta N_i$ is the number of atoms of \textit{i} added ($+1$) or removed ($-1$) at the defect. To avoid additional errors in the defect energy introduced by the equilibrium chemical potentials determined from the phase diagram, we used the energy of the pure elemental phases $\mu_i$ for this benchmark section. This choice does not affect the benchmark, but only shifts the value of the point defect energy.

Figure \ref{fig:surface_defect}(b) presents a comparison between uMLIP and DFT defect energies for 129 point defects across 32 chemical systems, including AlNi, \ch{CaSn3}, \ch{Cu3Au}, NaPb$_3$, NaAg$_4$, etc. Calculation details are listed in the \hyperref[sec:Methods]{Methods} section and the complete list of materials is provided in  \hyperref[sec:SI]{Supplementary} Table 2. Interestingly, the uMLIP calculated defect energies are mostly underestimated, similar to the trend observed in the surface energies in Fig. \ref{fig:surface_defect}(a).

\subsection{Solid-Solution Energetics}

Thermodynamic modeling of solubility in solid state systems such as metallic alloys \cite{VandeWalle2009} and high-entropy ceramics \cite{Lun2020_highEntropy} requires accurate energetics to capture the dependence of the energy on substitutional arrangements \cite{CE_2022, Zhong2023_PRXEnergy}. This dependence, relative to $k_BT$, determines the temperature scale at which mixing or order/disorder transitions occur \cite{Ceder_1993_Ising}. In this section, we use the mixing of Ca$^{2+}$ and Mg$^{2+}$ in the Ca$_x$Mg$_{2-x}$O$_2$ rocksalt as an example to examine the ability of uMLIPs to predict the behavior of the solid solution. The end members of the system, MgO and CaO are both rocksalts and the phase diagram has been previously studied both experimentally \cite{DOMAN_1963_CaMgO_experiment} and computationally \cite{CaMgO_2005_theory}.

We explore different possible Ca-Mg cation arrangements in the rocksalt at various CaO-MgO ratios and evaluate the corresponding energies (see \hyperref[sec:Methods]{Methods}). These 0K formation energies are shown in Fig. \ref{fig:surface_defect}(c), where each point corresponds to the energy of a specific Ca-Mg cation arrangement at a given Ca fraction. The predicted formation energies from all uMLIPs are positive, consistent with the low T immiscibility of CaO and MgO \cite{CaMgO_2005_theory}. We observe a systematic underprediction of the mixing energies and the energy difference between the uMLIPs and DFT at a specific Ca fraction. Among the uMLIPs, CHGNet's predictions closely approximate those of DFT, followed by those of M3GNet and MACE. We note that an underprediction of the formation energy would lead to an underestimation of the solubilization temperature in phase diagram calculations and an overestimation of the solubility limits at a given temperature \cite{Ceder_1993_Ising}.

\subsection{Phonon Properties}
\begin{figure*}[t]
\centering
\includegraphics[width=\linewidth]{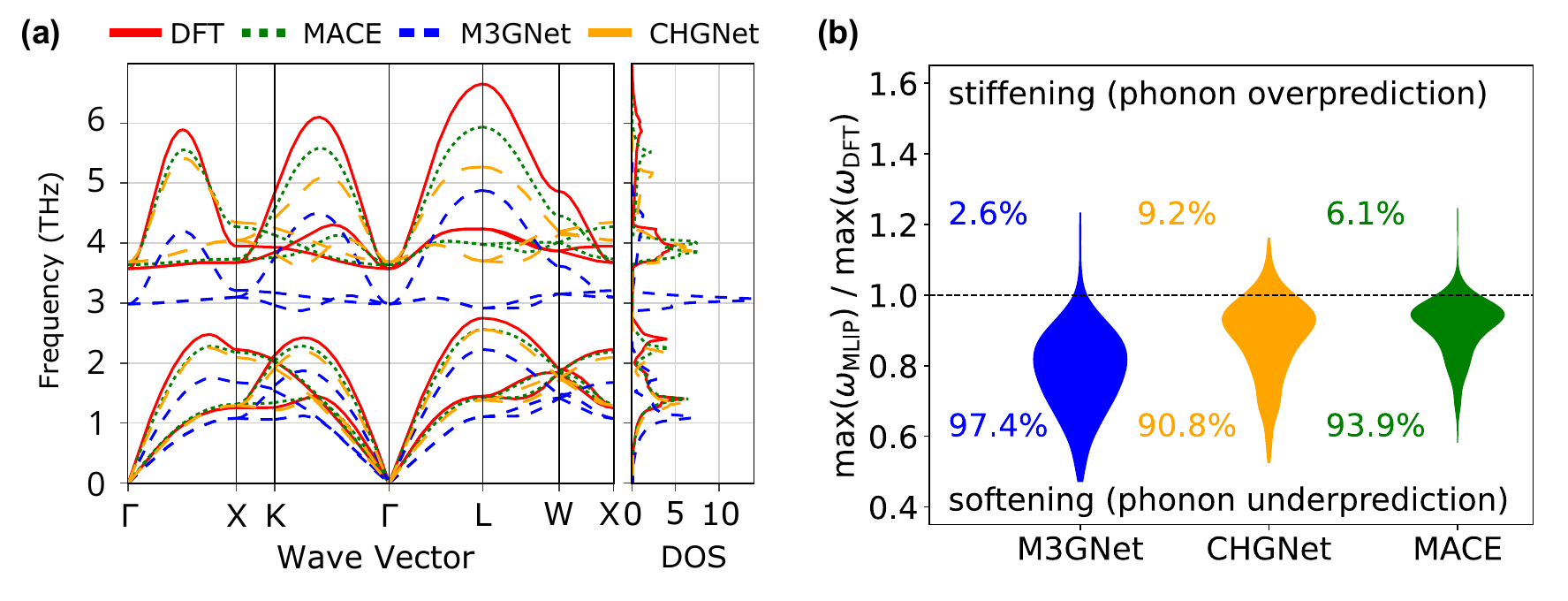}
\caption{\textbf{Softened phonon vibration modes in uMLIPs.} (a) the phonon dispersion relation and density of states (DOS) of Cs$_4$F$_4$(\texttt{mp-1784}) calculated with DFT and uMLIPs. Systematic underpredictions of phonon vibration frequencies are observed with all uMLIPs. (b) Distribution of ratios between uMLIP maximum frequency to DFT maximum frequency for 229 different compounds.}
\label{fig:phonon} 
\end{figure*}

Accurate descriptions of vibrational properties and phonon spectra are crucial for understanding a wide range of material characteristics, such as thermodynamic \cite{Walle_2002_phonon}, mechanical \cite{phonon_2015}, and thermal transport properties \cite{Yue_2020_phonon_thermal}. Predicting phonon frequencies represents a stringent test of the MLIPs' ability to capture the subtle energy and force landscape around equilibrium configurations. In this section, we benchmark the uMLIPs' performance on phonon frequencies by applying the finite displacement method \cite{Parlinski_1997_FDM} to calculate harmonic phonons.

Figure \ref{fig:phonon}(a) shows an example of uMLIP and DFT calculated phonon frequency on Cs$_4$F$_4$ (Materials Project ID \texttt{mp-1784}), where the solid red lines represent DFT phonon frequencies and the dashed lines show uMLIP phonon frequencies. While the overall shapes of the phonon bands are generally well-captured by the uMLIPs, a systematic reduction of the vibrational frequencies (i.e., the frequency magnitude difference of the branches at a given wave vector) is observed across all models compared to the DFT reference, particularly for the optical modes predicted by M3GNet (blue dashed line). The reduced vibrational frequency is an indication that the forces described by uMLIPs are systematically lower than the DFT values.

To quantify this softening behavior, we evaluate the ratio between the maximum phonon frequencies predicted by the uMLIPs and the corresponding DFT value for a diverse set of 229 materials (see \hyperref[sec:SI]{Supplementary} Table 3) from the PhononDB \cite{togo_firstprinciples_2023, togo_implementation_2023}. The distribution of these ratios is shown in Fig. \ref{fig:phonon}(b), which demonstrates that the majority ($>90\%$) of materials are found to be softened in uMLIPs compared to DFT, with the phonon frequency under-predicted. The result suggests that both the energy and force described by uMLIPs are softened for almost all chemical systems.

\subsection{Ion Migration Barriers}
\begin{figure*}[t]
\centering
\includegraphics[width=\linewidth]{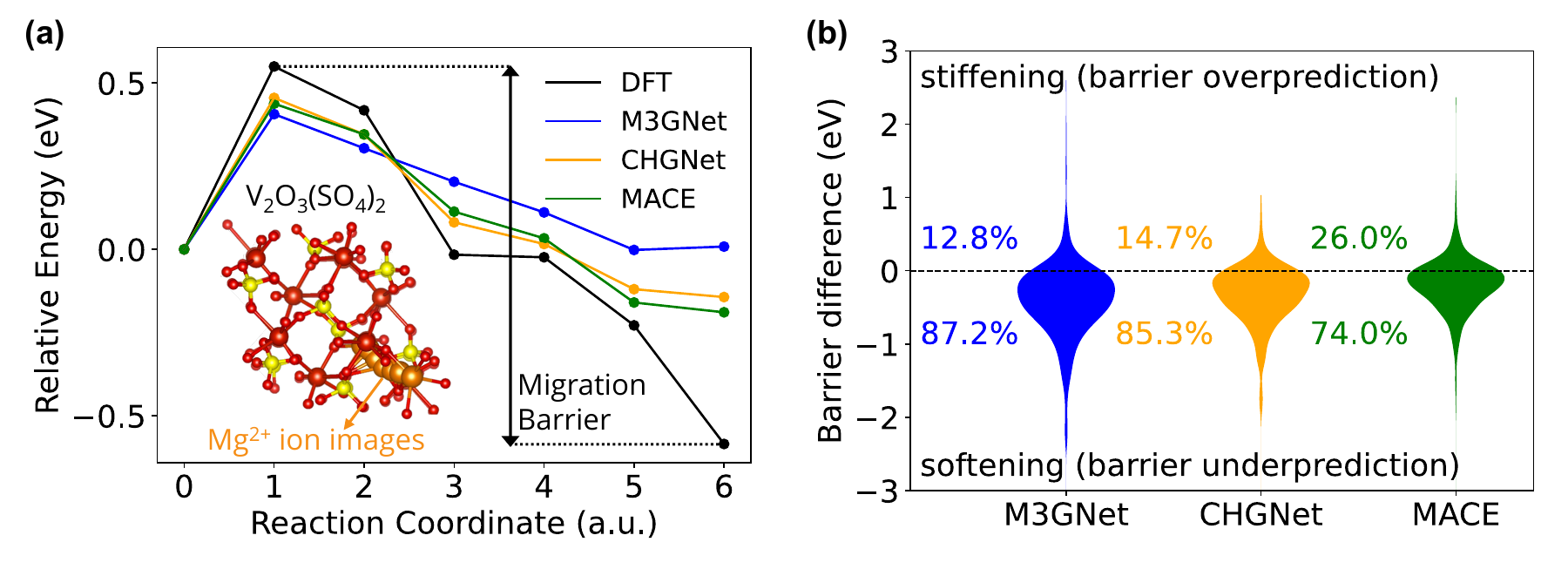}
\caption{\textbf{Underpredicted ion migration barriers in DFT and uMLIPs.} (a) An example of a Mg-ion migration path in \ch{V2O3(SO4)2} (\texttt{mp-28207}) with 6 intermediate images. The migration barrier is defined as the energy difference between the highest and lowest image. (b) The distribution of 477 energy barrier differences between uMLIPs and DFT, showing uMLIPs' tendency to underestimate the ion migration barriers.}
\label{fig:barrier} 
\end{figure*}

The migration barrier for an ion to move through a crystal structure forms the basis for evaluating the diffusion constants in a material and as such is critical to understand its functional or processing behavior. An accurate description of ion mobility is directly relevant in various applications, such as lithium-ion conductors for battery technologies \cite{NEB_2015}, and proton conductors for fuel cells \cite{Du_2020_proton_NEB}, etc. Because the migration barrier is determined by the extrapolation of the energy along the path between two stable sites, it is by definition also a poorly sampled configuration when uMLIPs are only fitted to local equilibrium configurations.

We employ uMLIPs and DFT to conduct a comprehensive assessment of 470 Mg-ion migration pathways in 110 distinct structures including oxides, halides, and sulfides \cite{Rutt_2022_NEB}. For all ion migration paths, we generate an initial guess of the minimum energy pathway based on the DFT charge density \cite{Shen_2020_chargedensity} and subsequently evaluate it with the approximate nudged elastic band (ApproxNEB) method \cite{Rong_2016_approxNEB}(see method section). ApproxNEB is different from regular NEB in that it does not perform a relaxation of the pathway but solely evaluates the energy along the predefined trajectory \cite{Rong_2016_approxNEB}. Figure \ref{fig:barrier}(a) presents the energy landscape of one Mg ion migration path in \ch{V2O3(SO4)2} (\texttt{mp-28207}), where the energies of each image have been referenced against the energy of image 0. The migration barrier is defined as the energy difference between the highest and lowest energies along the reaction coordinate. While all three uMLIPs are shown to capture the overall shape of the DFT energy along the path, we observe systematic energy under-predictions of uMLIPs resulting in under-predictions of migration barriers. MACE achieves the best performance with a 0.30 eV MAE against DFT, followed by CHGNet (0.39 eV) and M3GNet (0.49 eV). The parity plot of uMLIP barriers \textit{vs.} DFT barriers is provided in \hyperref[sec:SI]{Supplementary} Fig. 1 and shows that the majority of uMLIP barriers are under-predicted, similar to the result of the surface and defect benchmarks. Figure \ref{fig:barrier}(b) presents the distribution of the energy barrier difference between uMLIPs and DFT, from which we observe that all three uMLIPs show negative shifts in barrier predictions.

\subsection{PES softening scale for high-energy states}
\begin{figure*}[t]
\centering
\includegraphics[width=\linewidth]{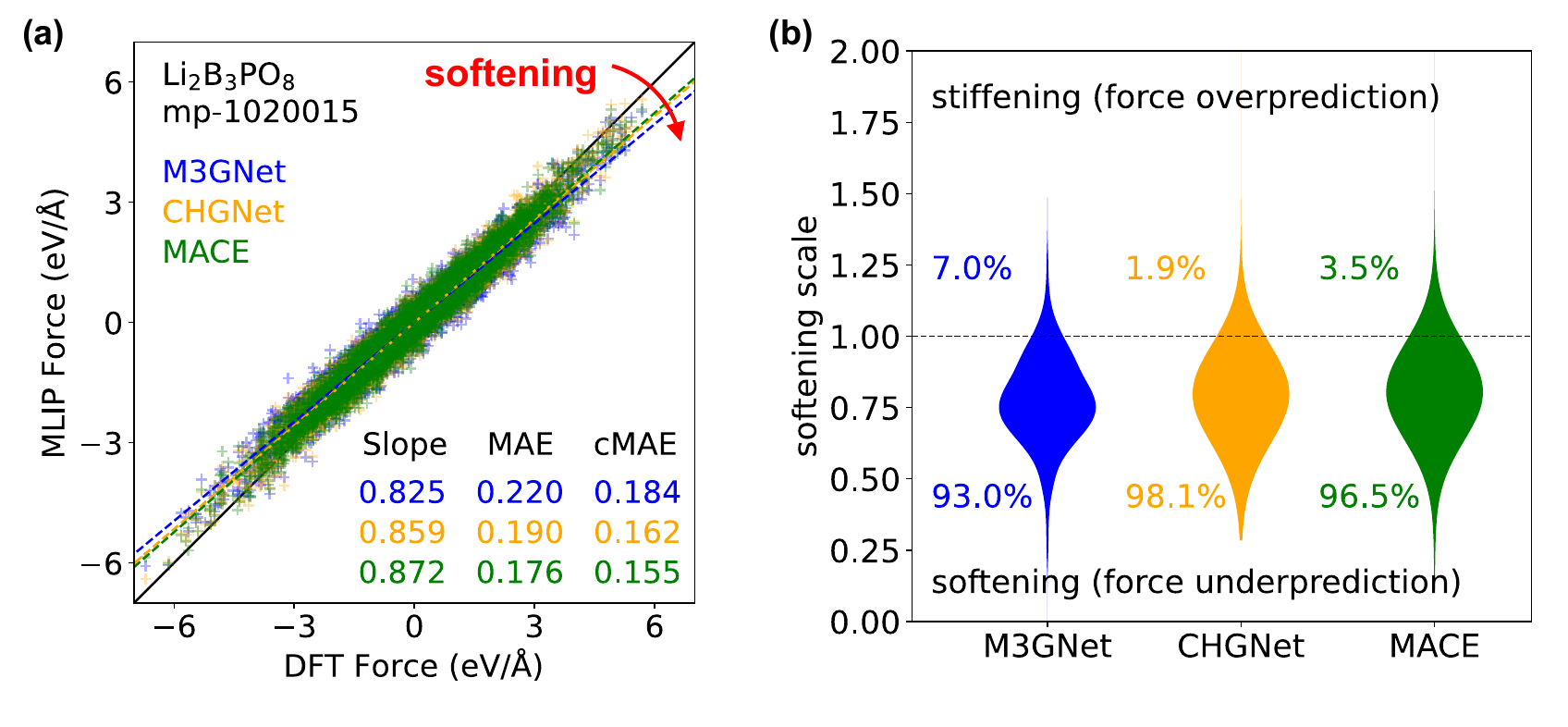}
\caption{\textbf{The PES softening scale from shifted force predictions.} (a) uMLIP forces \textit{vs.} DFT forces in high-energy states, sampled from high-temperature MDs of $\ch{Li2B3PO8}$(\texttt{mp-1020015}). Systematic softening of PES is indicated by the tilted distribution of forces from the diagonal. The softening scale is defined as the slope of the distribution, where softening is indicated by slope $<$ 1 . cMAE stands for corrected mean absolute error, which is the MAE if the softening scale is corrected to 1, equivalent to having the force distribution rotated back to diagonal. (b) Distribution of softening scales of 1000 compounds sampled from the WBM dataset, showing the PES softening behavior is universal across various chemical systems.}
\label{fig:off-eq} 
\end{figure*}

By definition, a machine learning model with only random errors should have its prediction error distribution centered at 0. However, all three uMLIPs are shown to not satisfy such criterion in our OOD benchmarks. The negative distribution shifts in Fig. \ref{fig:phonon}(b) and Fig. \ref{fig:barrier}(b) indicate the existence of systematic prediction errors in the uMLIPs. These systematic underpredictions of energies and forces act as a softening effect of the PES, creating systematic errors in the calculation of important material quantities. To quantify the magnitude of softening observed in our benchmark tasks, we define a softening scale parameter, which is calculated as the linear fitted slope of uMLIPs \textit{vs.} DFT forces in a material. As an example, Figure \ref{fig:off-eq}(a) shows an exemplary parity plot of uMLIPs \textit{vs.} DFT forces from sampled high-energy OOD atomic configurations of Li$_2$B$_3$PO$_8$ (Materials Project ID \texttt{mp-1020015}). These OOD atomic configurations are sampled away from the energy minimum in the PES, by applying high-temperature molecular dynamics (MD) simulations(see method section). The corresponding forces of each sampled state are subsequently evaluated using static calculations with uMLIPs and DFT.

The systematic PES softening effect shows up in Fig. \ref{fig:off-eq}(a) by the clockwise tilting of the distribution away from the diagonal. The slope of this distribution, extractable by linear regression, can be defined to be the PES \emph{softening scale}. In Fig. \ref{fig:off-eq}(a), we provide the fitted slopes and force MAEs of the three uMLIPs. When the softening scale is 1, the MLIP's force distribution aligns with the diagonal, indicating that the curvature of the MLIP-PES systematically agrees with DFT with only random errors present. A softening scale smaller than 1.00, which is observed for all the benchmark tasks, indicates a systematic underprediction of energy and forces that leads to an overall smoother PES curvature as illustrated in Fig. \ref{fig:pes}.

To investigate how broadly across chemistry the PES softening occurs, we collected 1000 different compounds from the WBM materials dataset by \citet{WBM_2021}, which was generated by elemental substitution of Materials Project compounds and therefore contains only crystalline structures that are not included in the pre-training dataset of the three uMLIPs. For each of these compounds, 10 high-energy states away from the PES energy minimum are sampled with a 1000K MD simulation, and the softening scale is extracted from a linear fit with uMLIP and DFT forces. Figure \ref{fig:off-eq}(b) presents the distribution of the PES softening scale for these 1000 WBM compounds, and shows that for the majority ($>90\%$) of the compounds, the softening scale is smaller than 1 for all 3 uMLIPs we have tested. This result indicates the systematic softening behavior is universal across all chemical systems in current uMLIP models.

\subsection{Data-efficient fine-tuning}
\begin{figure*}[t]
\centering
\includegraphics[width=\linewidth]{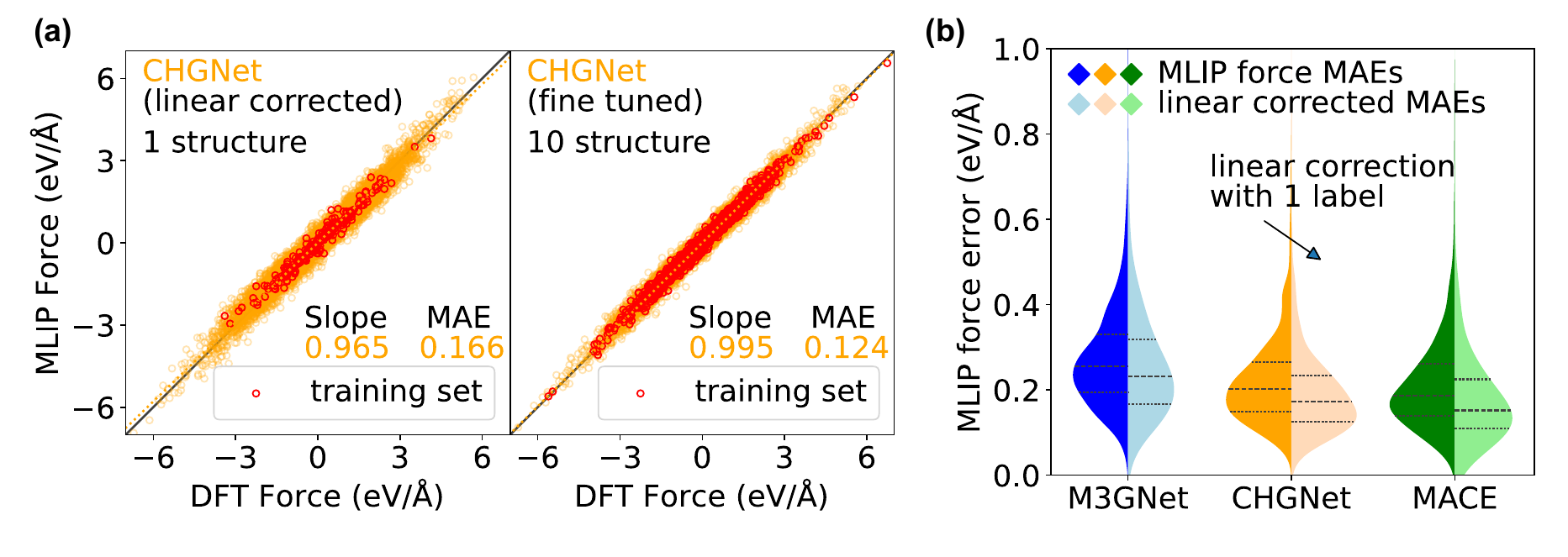}
\caption{\textbf{Linear correction with one label solves PES softening.} (a) Parity plots of fine-tuned CHGNet predictions on $\ch{Li2B3PO8}$ (\texttt{mp-1020015}), with the training force labels plotted in red and pre-excluded test force labels plotted in orange. Left: fine-tuned CHGNet with a linear correction and a single DFT label solves the softening issue and greatly reduces force MAE from 0.190 eV/$\mathrm{\AA}$ to 0.166 eV/$\mathrm{\AA}$. Right: a more realistic fine-tuning example that optimizes all model parameters with 10 DFT labels, which further decreases the force MAE. (b) Distribution of force MAEs and linear corrected MAEs (cMAEs) for 1000 WBM compounds, showing uMLIP force errors can be greatly reduced by fine-tuning with a single data point. Quartiles are labeled by dashed lines.}
\label{fig:linear-correction} 
\end{figure*}

The PES softening issue appears as a tilted distribution of forces in the parity plot as shown in Fig. \ref{fig:off-eq}(a). Intuitively, one can rotate the distribution back to the diagonal to reset the softening scale to 1 hereby reducing the prediction error. In this scenario, we define cMAE as the linearly corrected mean absolute error if the uMLIP force distributions were rotated back to align with the diagonal. As shown in Fig. \ref{fig:off-eq}(a), the cMAEs are considerably reduced from the original MAE from 0.220/0.190/0.176 eV/$\mathrm{\AA}$ to 0.184,0.162,0.155 eV/$\mathrm{\AA}$ for M3GNet/CHGNet/MACE, respectively. This observation suggests that a considerable fraction of force errors from uMLIP are likely to be systematic and can be easily corrected to reduce force errors. 

Mathematically, rotating the force distribution is equivalent to multiplying every force value by a scalar, which can be realized by multiplying the MLIP energy by a scalar term
\begin{equation}
\begin{aligned}
    E^\text{corr} &= c * \textbf{MLIP}(\{\vec{r}_i\}, \{C_i\}),\\
    f^\text{corr}_i &= -\frac{\partial{E^\text{corr}}}{\partial{\vec{r}_i}} = c * f_i.
\end{aligned}
\label{eq:fine-tune} 
\end{equation}
It is noted that the above formulation is equivalent to fine-tuning a MLIP by fixing all model weights except a scalar linear layer, which essentially modifies only the scalar parameter $c$ in Equation \eqref{eq:fine-tune}. Since only a scalar parameter requires modification, only one single label (1 force component) is needed for the training. In the left part of Fig. \ref{fig:linear-correction}(a), we show the result when pre-trained CHGNet is fine-tuned with an added hypothetical scalar linear layer (see \hyperref[sec:Methods]{Methods}), trained on only a \emph{single} high-energy configuration of $\ch{Li2B3PO8}$ (\texttt{mp-1020015}). The test forces, which originate from the same set of atomic arrangements in Fig. \ref{fig:off-eq}(a), are labeled in orange and the training forces from the single additional configuration are labeled in red. The linear corrected CHGNet exhibits a softening scale of 0.965 and a force MAE of 0.166 eV/$\mathrm{\AA}$, improved from 0.859 and 0.190 eV/$\mathrm{\AA}$ in the pre-trained CHGNet as shown in Fig. \ref{fig:off-eq}. The estimated cMAE is 0.162 eV/$\mathrm{\AA}$ when the softening is corrected to 1, which is close to the force MAE of 0.166 eV/$\mathrm{\AA}$ that is achieved by fine-tuning the scalar linear layer. Hence, a linear correction with one high-energy OOD configuration indeed operates as a rotation of the force distribution back to the diagonal, substantially eliminating the systematic softening error and considerably reducing the force MAE.

We propose that the cMAE derived from the linear correction serves as an approximate lower bound for the expected error reduction from fine-tuning uMLIPs. While a linear correction with one label stands as an extreme case, a typical fine-tuning process involves hundreds and thousands of structure labels that can further reduce the MAE of the model. We tested fine-tuning the pretrained CHGNet by optimizing all model parameters with 10 training structures, and the resulting force parity plot is shown on the right of Fig. \ref{fig:linear-correction}(a). Compared to the linear correction with only one configuration, the right panel in Fig. \ref{fig:linear-correction}(a) shows that a very small dataset of 10 training structures further reduces the MAE to 0.125 eV/$\mathrm{\AA}$. By statistically evaluating the distribution of force MAEs and cMAEs for the 1000 WBM structures, we present their fine-tuning error-reduction lower-bounds in Fig. \ref{fig:linear-correction}(b). From the observed distribution, considerable error reduction ($\sim15\%$) can be adequately achieved with a simple linear correction.

These results suggest a theoretical explanation for the commonly observed data-efficient performance boost that is achievable by fine-tuning foundational uMLIPs compared to training randomly initialized MLIPs. The data efficiency arises from the observation that a significant part of the MAEs in pre-trained uMLIPs are highly systematic, which can be efficiently amended by optimizing a fraction of model parameters with a small amount of data. In \hyperref[sec:SI]{Supplementary} Fig. 2, we present a comparison between the force error in the fine-tuned CHGNet models to those trained from scratch. The result demonstrates that the data-efficient error reduction is a unique advantage that only applies to foundational uMLIPs that have been well-pre-trained, and randomly initialized MLIPs will require substantially more data points to converge to the same performance.

\section{Discussion}
The design and discovery of novel materials raises the need for advanced simulation tools capable of efficiently and accurately describing the intricate details of atomic interactions. MLIPs offer a potential solution to bridge the gap between quantum mechanical accuracy and affordable computation cost by learning and emulating complex atomic interactions. Recent work on pre-training foundational MLIPs with comprehensive material datasets has opened up the possibility for out-of-box use of robust universal interatomic potentials \cite{M3GNet_2022, CHGNet_2023, GNOME_2023, MACEMP_2023, Zhang_2024_DPA1}.

Unlike DFT, MLIPs cannot by default be expected to perform well in a configurational space where they have not been trained. We therefore benchmark the performance of three uMLIPs for multiple OOD modeling tasks including surfaces, defects, solid-solution energetics, phonon vibration modes, ion migration barriers, and more general high energy states. These states are under-represented in the widely-used pre-training dataset \cite{MaterialProject, M3GNet_2022, CHGNet_2023} that only consists of bulk crystalline materials. For the properties tested in this work, we observe a universal softening of the PES, characterized by the uMLIPs' under-prediction of energies and forces. 

The uMLIP datasets are primarily drawn from Materials Project \cite{MaterialProject} ionic relaxation trajectories and are therefore largely distributed around the energy minima of the PES. Consequently, the uMLIPs are exposed to a limited range of atomic configurations and force gradients, leading to difficulties in accurately capturing the energy landscapes and steep gradients associated with OOD states and processes like ion migrations and phase transformations. 

We found similar signs of softening in the published literature, though less attention was dedicated to an in-depth examination of the softening issue. \citet{Pandey_2021} and \citet{Bartel_2021} presented an extrapolation issue arising from a distribution shift when training a CGCNN \cite{Xie_Grossman_2018} energy predictor with ICSD data \cite{Belsky_2002_ICSD}. The CGCNN model trained with only experimental stable materials experienced a six-fold increased prediction MAE when applied to hypothetical crystal structures in the Materials Project \cite{MaterialProject}. Furthermore, the Google DeepMind's GNoME uMLIP exhibited pronounced softening tendencies when trained on the M3GNet dataset \cite{M3GNet_2022}, as evidenced in the Supplementary Information of Ref. \cite{GNOME_2023}, similar to our observation in Fig. \ref{fig:off-eq}(a). After being trained on the expanded dataset of 89 million structures, the softening issue in GNoME was shown to be mitigated but not fully eliminated, which is shown in the Supplementary Figs. S34--S37 from Ref. \cite{GNOME_2023}. These examples underscore the universality of the PES softening issue across various models and datasets, highlighting the importance of the systematic benchmark and analysis undertaken by our study to address this challenge.

The observed limitations of current uMLIPs raise questions about the effect of model size and expressive capacity on their ability to capture the intricate details of the PES \cite{Ko_Ong_2023}. The MACE model with 4.69 Million parameters, which is around 11 times the size of the CHGNet and 21 times the size of M3GNet, shows improved MAE and decreased softening compared to the smaller uMLIPs. The better performance of larger uMLIPs aligns with the previous study by \citet{NNScaling_2023} on the scaling of model performance as a function of MLIP capacity. The observed relationship between model capacity and performance prompts further inquiry into the extent to which the parameter size of current uMLIPs influences the PES softening issue, and whether the softening can be minimized by scaling to a larger, yet reasonable model size without expanding the dataset. In \hyperref[sec:SI]{Supplementary} Fig. 4 and 5, we show the distribution of softening scale and force MAEs for two additional uMLIPs: \texttt{CHGNet-matgl} and \texttt{M3GNet-matgl}, which were also pre-trained using Materials Project database. The \texttt{CHGNet-matgl} with increased model size and \texttt{M3GNet-matgl} with enhanced sampling \cite{Qi_sampling_2023} demonstrate decreased softening effect and improved force predictions. While the scope of current work does not explicitly investigate the effect of model size, further studies could provide a better explanation of the number of model parameters needed to describe a universal potential energy surface.

Fortunately, we demonstrate the softening issue can be effectively resolved by including a minimal amount of high-energy OOD training points in fine-tuning. Our result not only provides a guideline to avoid softening issues when applying uMLIPs to atomic modeling, but more importantly, derives an explanation for the frequently observed data-efficient fine-tuning of foundational MLIPs. Our result suggests that a significant portion of the MAE in uMLIPs is highly systematic and therefore can be efficiently corrected by a small amount of data. In addition to the robustness of uMLIPs that has been acknowledged as an advantage obtained from pre-training \cite{CHGNet_2023, GNOME_2023}, our study elucidates another benefit of fine-tuning foundational MLIPs -- the data-efficient systematic error correction that is unavailable for training a randomly initialized MLIP. Our study serves as a guideline for researchers attempting to fit interatomic potentials for their systems of interest.

In summary, our work presents an in-depth analysis of the softening effect of uMLIPs observed in a series of OOD materials benchmarks, from which we provide guidelines for the fine-tuning effects of uMLIPs. With the observed limitation of current uMLIPs, we advocate the need for an improved next-generation dataset for training foundational atomic models, and more investigation in the role of model complexity. Despite significant efforts dedicated to model design and training strategies, less emphasis has been placed on constructing comprehensive and well-curated open-source materials datasets. The current foundational models still rely on datasets that were not originally generated for machine learning purposes. Apart from diversifying the chemical space, our findings highlight the importance of ensuring a comprehensive sampling of the PES in generating a reliable MLIP dataset. We believe a next-generation foundational atomic dataset with improved sampling will be pivotal for the development of MLIP and atomistic simulations.

\section{Acknowledgments}
This work was funded by the U.S. Department of Energy, Office of Science, Office of Basic Energy Sciences, Materials Sciences and Engineering Division under Contract No. DE-AC0205CH11231 (Materials Project program KC23MP). The work was also supported by the computational resources provided by the Extreme Science and Engineering Discovery Environment (XSEDE), supported by National Science Foundation grant number ACI1053575; the National Energy Research Scientific Computing Center (NERSC), a U.S. Department of Energy Office of Science User Facility located at Lawrence Berkeley National Laboratory; and the Swift Cluster resource provided by the National Renewable Energy Laboratory (NREL). The authors would also like to thank Tsz Wai Ko and Yuanqi Du for helpful discussions.

\section{Methods}
\label{sec:Methods}
\subsection{uMLIP versions}
The table below shows the details and versions of the uMLIPs tested.
\begin{table}[ht]
\centering
\begin{tabular}{c|c|c|c|c}
\hline\hline
Model                    & Version        & ModelSize & DataSet                  & DataSize \\ \hline
M3GNet \cite{M3GNet_2022} & \href{https://github.com/materialsvirtuallab/m3gnet/tree/main/pretrained/MP-2021.2.8-EFS}{2021.2.8}   
& 227.5K      & MPF \cite{M3GNet_2022}    & 188.3K \\
CHGNet \cite{CHGNet_2023} & \href{https://github.com/CederGroupHub/chgnet/tree/main/chgnet/pretrained/0.3.0}{v0.3.0}       
& 412.5K      & MPtrj \cite{CHGNet_2023}  & 1.58M  \\
MACE \cite{MACEMP_2023}   & \href{https://github.com/ACEsuit/mace-mp/blob/main/mace_mp_0/2023-12-03-mace-128-L1.sh}{2023.12.03}     & 4.69M       & MPtrj                    & 1.58M  \\
\hline\hline
\end{tabular}
\caption{uMLIP Model Specifications}
\label{table:model_specifications}
\end{table}

\subsection{Materials Modeling Tasks}

For surface energy calculations, stoichiometric and symmetric slabs are generated with up to a maximum Miller index of 2 in three directions. Minimum slab thickness of 10 $\mathrm{\AA}$ and minimum vacuum length of 10 $\mathrm{\AA}$ are used for DFT to ensure convergence of surface energy \cite{Sun_2013_surface}. When relaxing the slab, in-plane lattice vectors are fixed to their bulk value. The ionic relaxations are converged to a maximum interatomic force criteria of 0.05 eV/$\mathrm{\AA}$ for all uMLIPs.

For defect energy calculations, defects in elemental phases as well as binary metallic compounds are considered. The defect structures are fully relaxed and referenced to the bulk energy. The off-stoichiometric defect energies (ex: vacancy defect) are referenced to the chemical potential of the pure elemental phase, instead of any chemical potential corresponding to multi-phase equilibria in the phase diagram. This is done deliberately to avoid additional errors associated with calculating the phase diagram using the uMLIPs. For all uMLIPs, the ionic relaxations are converged while the a maximum interatomic force is 0.05 eV/$\mathrm{\AA}$.

For solid-solution calculation in Ca$_x$Mg$_{2-x}$O$_2$, we randomly select different Ca-Mg orderings (up to 52 number of configuration) at each Ca concentration and evaluate the energy of the configuration with ionic relaxation with DFT or uMLIPs.

For phonon calculations, we use the \texttt{phonopy} workflow as implemented in \texttt{atomate2} \cite{ganose_atomate2_2024} with relaxation convergence and supercell settings identical to those used in \citet{MACEMP_2023}. The DFT referenced data are taken from the PhononDB \cite{togo_firstprinciples_2023, togo_implementation_2023}. We restrict benchmarking materials without magnetism and U-corrections. Moreover, we removed the non-analytic corrections (NAC) to PBE phonons which are derived from the Born effective charges as these are unavailable from uMLIPs which have no concept of electronic structure. In practice, a future hybrid uMLIP-DFT workflow could perform a single DFT static at the uMLIP relaxed structure to obtain Born charges and post-hoc apply non-analytic corrections to the uMLIP phonon spectrum. However, such a hybrid workflow while necessary in practice, would not affect the results of this benchmark concerned specifically with the ML-obtainable parts of the spectrum.

The ion migration barrier DFT data are collected from the work of \citet{Rutt_2022_NEB}, in which the ApproxNEB algorithm \cite{Rong_2016_approxNEB} was used to evaluate $Mg^{2+}$ ion migration barriers. The key difference between ApproxNEB with regular NEB \cite{Henkelman_2000_NEB} is that ApproxNEB relaxes each image along the migration path independently, while NEB relaxes the migration path collectively. In the ApproxNEB method, an initial guess of the ion migration path is interpolated based on the charge density of the host structure. The energies associated with suggested image structures are calculated by the constrained relaxation that fixes the moving ion and lattice vectors. The ApproxNEB method was shown to provide a comparable barrier within 20 meV error of NEB and reduce the computational time significantly for materials where the path is not too complex\cite{Rong_2016_approxNEB}.

To sample the high-energy states, we randomly select 1000 structures from the WBM dataset \cite{WBM_2021}. For each structure selected, a 20 ps ,1000K molecular dynamics simulation is performed under constant number of particles, volume, and temperature (NVT) ensemble with the pre-trained CHGNet, and 10 structures are subsequently sampled from each MD trajectory \cite{Qi_sampling_2023}. $+3\%$ strain and a $-3\%$ strain are applied along three lattice dimensions for 4 out of the 10 structures to sample strained configurations. All the force MAEs and fine-tuning are calculated with the three-dimensional force components rather than the absolute magnitude of forces.

\subsection{Fine-tuning}

For the fine-tuning of CHGNet uMLIP, the models are trained with energy, force, and stress labels with 0.1-100-0.1 loss fractions under the mean squared error (MSE) loss criterion. The structures and labels are taken from a DFT ab-initio MD trajectory data of $\ch{Li2B3PO8}$ (\texttt{mp-1020015}), where 100 structures are reserved for the test set, as shown by the orange points in Fig. \ref{fig:linear-correction}(a). The train-validation ratio is set to be 9:1. As a result, 9 out of the 10 training structures in the right panel of Fig. \ref{fig:linear-correction}(a) are actually used for gradient back-propagations. The \texttt{Adam} optimizer \cite{Kingma_2014_ADAM} is used with a learning rate of 1e-3 that cosine decays to 1e-5 in 100 epochs. The model checkpoint of best validation force MAE is collected for test set predictions. For the model trained with only 1 structure, the last-epoch checkpoint is used instead. 

The linear correction of CHGNet is realized by adding a hypothetical scalar linear before the energy prediction. The weight of the scalar linear layer is initialized to be 1, therefore not influencing the energy prediction before being optimized. During the linear correction, all CHGNet model parameters are frozen except for the added scalar linear layer.

\subsection{DFT calculations}
DFT calculations were performed with the \textit{Vienna ab initio simulation package} (VASP) using the projector-augmented wave method \cite{kresse1996VASP, kresse1999PAW}. All calculation settings are generated using \texttt{pymatgen} \texttt{MPRelaxSet} to ensure all DFT results are compatible with Materials Project DFT calculations \cite{pymatgen_2013}. All the calculations were converged to at least $10^{-5}$ eV in total energy for electronic steps and 0.02 eV/Å in interatomic forces for ionic steps.

\section{Supplementary Information}
\label{sec:SI}
The supplementary material is available at URL-When-Published.

\bibliography{references}

\end{document}